\documentclass[aps,floats,groupedaddress,twocolumn]{revtex4}
\usepackage[dvips]{graphics}
\newcommand\hbarr{\hbar}
\newcommand\citen{\cite}

\newcommand\Eq[1]{Eq.~(\ref{eq:#1})}
\newcommand\Fig[1]{Fig.~\ref{fig:#1}}
\newcommand\Ref[1]{Ref.~\citen{#1}}

\def\oo/{\discretionary{o-}{o}{o\"o}}

\newcommand\kB{k_{\mbox{\tiny B}}}
\newcommand\hw{\hbarr\omega}
\newcommand\hwt{{1\over2}\hbarr\omega}
\newcommand\xO{x_{0}}

\newcommand\ennt{\tilde{n}}
\newcommand\enet{\tilde{n}}

\newcommand\mut{\tilde\mu}
\newcommand\gammat{\tilde{\gamma}}
\newcommand\lapt{\tilde\nabla^{2}}
\newcommand\psit{\tilde\psi}
\newcommand\Phit{\tilde\Phi}
\newcommand\epsit{\tilde\epsilon}

\newcommand\Lambdat{\tilde\Lambda}
\newcommand\eLe{\Lambdat + 2\gammat\ennt}
\newcommand\Zeff{(x^2+2\gammat\enet-\mut)}

\newcommand\dV[2]{d^#1\!#2\,}

\newcommand\eHFB{\epsit_{\mbox{\tiny HFB}}}
\newcommand\eHF{\epsit_{\mbox{\tiny HF}}}

\def\EQQ{\!\!&=&\!\!}
\def\EQQh{\!\!&\hphantom{=}&\!\!}

\newcommand\gee[2]{\ensuremath{g_{#1}(e^{#2})}}

\newcommand\xx{{\mathbf x}}

\newcommand\gradt{\tilde{\mbox{\boldmath$\nabla$}}}
\newcommand\kk{\mbox{\boldmath$\kappa$}}

\newcommand\Psid{\Psi^{\dagger}}

\begin{document}

\title{The Two-Dimensional Bose-Einstein Condensate}

\author{Juan Pablo Fern\'andez and William J. Mullin}
\affiliation{Department of Physics, University of Massachusetts,
		Amherst, MA 01003, U.S.A.}

\date{\today}

\begin{abstract}
We study the Hartree-Fock-Bogoliubov mean-field theory as applied to a
two-dimensional finite trapped Bose gas at low temperatures and find
that, in the Hartree-Fock approximation, the system can be described
either with or without the presence of a condensate; this is true in
the thermodynamic limit as well.  
Of the two solutions, the one that includes a condensate has a lower
free energy at all temperatures.  However, the Hartree-Fock scheme
neglects the presence of phonons within the system, and when we allow
for the possibility of phonons we are unable to find condensed
solutions; the uncondensed solutions, on the other hand, are valid
also in the latter, more general scheme.  Our results confirm
that low-energy phonons destabilize the two-dimensional condensate.
%
%
\end{abstract}

\maketitle

\section{Introduction}

In a recent paper~\cite{gorlitz}, one of the groups that pioneered the
formation and detection of Bose-Einstein condensation (BEC) in
harmonically trapped atomic gases~\cite{Wieman,Ketterle,Hulet} reports
the creation of \hbox{(pseudo-)} two-dimensional condensates.  These
have been produced by taking a three-dimensional condensate of
${}^{23}$Na~atoms and carrying out two independent processes on it:
1)~Initially, one of the confining frequencies (that in the $z$
direction, $\omega_z$, in order to minimize the effects of gravity) is
increased until the condensate radius in that dimension is smaller
than the healing length associated with the interaction between atoms
(taken here to be repulsive and parametrized by the two-body
scattering length~$a$).  This is not sufficient to reduce the
dimensionality, however, since the atoms, each of which has mass~$m$,
will literally squeeze into the third dimension if there are more than
\begin{equation}
	N = \sqrt{\frac{\mathstrut32\hbar}{\mathstrut225ma^2}}\,
		\sqrt{\frac{\omega_z^3}{\omega_x^2\omega_y^2}}
\end{equation}
of them in the trap.  
2)~Consequently, the number of atoms in the condensate must be
reduced; this is achieved by exposing the condensate to a thermal
beam.  The reduction in effective dimensionality becomes apparent when
the aspect ratio of the condensate, which is independent of~$N$ in~3D,
starts to change as the number of atoms is gradually reduced.  The
condensates thus produced have a number of atoms that ranges between
$10^4$~and~$10^5$.

This constitutes an important experimental contribution to the
long-standing debate about the existence of BEC in two dimensions.  It
is a well-known fact (and a standard textbook exercise) that BEC
cannot happen in a 2D homogeneous ideal gas; a rigorous mathematical
theorem~\cite{HT} extends this result to the case where there are
interactions between the bosons.  When the system is in a harmonic
trap, on the other hand, BEC can occur in two
dimensions~\cite{bagklep2} below a temperature~$\kB T_c =
\hbar\omega\sqrt{6N/\pi^2}$, but the theorem is once again valid when
interactions between the bosons are considered~\cite{mulLong}: while
there is a BEC in the 2D system, it occurs at~$T=0$.

The preceding discussion is valid only in the thermodynamic limit,
which in the particular case of an isotropically trapped system
consists of making $N\to\infty$ and $\omega\to0$ in such a way that
$N\omega^2$ remains finite~\cite{Damle,mulLong}.  The question remains
whether a phenomenon resembling BEC---that is, the accumulation of a
macroscopic number of particles in a single quantum state---occurs or
not when the system consists of a \emph{finite} number of particles
confined by a trap of finite frequency, as is certainly the case in
experimental situations.  If there is such a phenomenon, one would
like to know more about the process by which this ``condensate'' is
destabilized at finite temperatures as $N$~grows.

Some authors~\cite{kagan87,kagan00} have considered, in the finite
homogeneous 2D case, the possibility of a BEC. A similar
analysis~\cite{petrov} was considered for a quasi-2D trapped gas; the
latter reference finds that the phase fluctuations in the condensate
vary with temperature and particle number as
\begin{equation} 
        \langle(\delta\phi)^2\rangle \propto T\log N 
\end{equation}
which diverges for finite temperature as~$N\to\infty$, as one would
expect.  For finite~$N$, on the other hand, the fluctuations are
tempered at very low temperature; since the coherence length, though
finite, is still larger than the characteristic length imposed on the
system by the trap~\cite{filshev} or by walls, one can speak of a
``quasicondensate.''  It is this quasicondensation that we wish to
study in this paper.  Reference~\citen{jakkola} reports the
observation of a quasicondensate in a homogeneous system of atomic
hydrogen adsorbed on~${}^4$He; at this point we cannot tell if the
condensates reported in \Ref{gorlitz} are actually quasicondensates.

One can approach the study of the 2D Bose gas by employing mean-field
theory, which in this context refers to the Hartree-Fock-Bogoliubov
(HFB)~equations and various simplifications thereof that we will
review in quantitative detail and classify in the present work.  The
HFB~theory has been remarkably successful in treating the 3D~case;
though we give a few representative references below, we refer the
reader to \Ref{RMPBEC}, which includes a comprehensive review and an
exhaustive list of references, and will henceforth concentrate on the
work that has been carried out in 2D.

The HFB theory assumes from the outset that the system is partially
condensed, and proposes separate equations to describe the condensate
and the uncondensed (thermal) component.  The condensate is described
by a macroscopic wavefunction that obeys a generalized
Gross-Pitaevski\u\i{}~(GP) equation~\cite{Gross,Pita}, while the
noncondensate consists of a superposition of Bogoliubov quasiparticle
and ``quasihole'' excitations weighted by a Bose distribution.  The
expression for the noncondensate can be simplified by neglecting the
quasihole excitations (the Hartree-Fock
scheme)~\cite{HuseSiggia,BagKlep}, by performing a semiclassical WKB
approximation~\cite{VdB}, or by combining both of
these~\cite{GPSTT,GPS,HKN}.  Each one of the above schemes can be
further simplified, when the thermodynamic limit is approached, by
neglecting the kinetic energy of the condensate: this is the
Thomas-Fermi limit~\cite{GSL,BaymPeth}.  Finally one can neglect the
interactions between thermal atoms, arriving at the semi-ideal
model~\cite{Stamper,Minguzzi}.

The semi-ideal model has been implemented in 2D~\cite{Turks,kim1}, but
has been found to yield unphysical results as the interaction strength
becomes sizable.  The full-blown HFB~model does not appear to be much
more successful, at least when considered in the semiclassical limit.
In a previous paper~\cite{mulShort} one of us found that, below a
certain temperature, the introduction of a condensate in the
Thomas-Fermi limit (corresponding to the thermodynamic limit) renders
the HFB~equations incompatible, with the noncondensate density
becoming infinite at every point in space.  The singularity occurs at
the low end of the energy spectrum, indicating that the condensate is
being destabilized by long-wavelength phonons; this interpretation in
terms of phase fluctuations had already been proposed for the
homogeneous system~\cite{HT,Reatto}.  In this work we also report our
inability to find self-consistent semiclassical solutions to the
HFB~model when a finite trapped system is considered.

Moreover, it was recently discovered ~\cite{bhaduri,brandon} that it
is possible, in the 2D case exclusively, to solve the HFB~equations
semiclassically at \emph{any} temperature without even having to
invoke the presence of a condensate (thus obtaining what we will call
an ``uncondensed'' solution).  In other words, it is possible to
simply cross out the condensate component and solve for the system to
a temperature close to~$T=0$.  The solution thus obtained shows an
accumulation of atoms at the center of the trap and yields a bulge in
density similar to that caused by the presence of a condensate, even
though no state is macroscopically occupied.

These results appear to reinforce the conclusion that BEC~cannot
happen in the two-dimensional trapped system.  However, we are still
confronted with the experimental results described above.
Furthermore, two different Monte Carlo
simulations~\cite{Heinrichs,Pearson} show significant concentrations
of particles in the lowest energy state for finite~$N$, though it must
be said that this method provides little information about the types
of excitations that contribute to the disappearance of the condensate,
and that it is difficult to carry out such simulations on very large
systems.

The HFB~model cannot be discounted at this stage either: when we
restrict ourselves to the Hartree-Fock approximation, it is possible
to find self-consistent solutions (henceforth referred to as
``condensed'') involving a condensate, both for finite systems and in
the thermodynamic limit, when using either the discrete set of
equations~\cite{kim2} or the WKB~approximation~\cite{Ibiza} to treat
the noncondensate.  The Hartree-Fock approximation neglects phononlike
excitations, so it is not surprising that it yields solutions.
However, one may be able to justify its usefulness.

It is known that, in the infinite homogeneous system, infrared
singularities occur but are renormalized by interactions, providing,
in essence, a cutoff at a low wavenumber $k_0\approx
\sqrt{nmU}$~\cite{fishhoh,prokofio,popovbook}, where $n$ is the
density, $m$ the particle mass, and $U$ the effective interaction
strength.  Indeed, it is possible to estimate the
Berezinskii-Kosterlitz-Thouless~(BKT) transition temperature by simply
cutting off the ideal-gas density expression at this
$k_0$~\cite{prokofio}.  Presumably a similar situation occurs in the
trapped case.  The Hartree-Fock approach provides a convergent theory
by cutting off the singularities at a wavenumber similar to that of
more rigorous theories.  Whether such an approach gives a reasonable
estimate of the BKT~transition temperature, the superfluid density, or
a quasicondensate density for the finite system will need to await a
more rigorous theoretical approach to the interacting 2D~trapped
gas~\cite{Stoofnew}.

Given the above limitations, we analyze the character of the BEC in
the 2D trapped system by solving the coupled equations of the theory. 
We find that, in the Hartree-Fock scheme, it is possible to find both
condensed and uncondensed solutions for the two-dimensional equations. 
We also calculate the free energy corresponding to each one and find
that the condensed solution has a lower free energy at all
temperatures, which appears to imply that, at least at this level of
approximation, the uncondensed solution is unphysical or metastable. 
The condensed solution will be ``preferred'' over the
uncondensed one, and we assume that the solution represents an 
approximation to a quasi-condensate.

It is evident from our discussion that our approach to BEC in 2D
trapped systems is a preliminary one and that a further analysis that
takes fully into account the BKT~transition is necessary.  A start to
answering this need has been presented in \Ref{Stoofnew}, and we
intend to return to this problem ourselves in the future.

\section{The Model} \label{sec:Model}

\subsection{The Hartree-Fock-Bogoliubov equations}

Throughout this paper we use a dimensionless system of units in which
all lengths are scaled by the oscillator length
$\xO\equiv\sqrt{\hbar/m\omega}$ and all energies are expressed in
terms of the one-dimensional ground-state energy of the oscillator,
$\hw/2$.  Dimensionless variables will in general carry a tilde: for
example, the total density~$n$ becomes~$\enet\equiv\xO^2 n$ and the
chemical potential~$\mut\equiv\mu/\hwt$.

The HFB equations~\cite{RMPBEC,GPS,Fetter72,Griffin96} result from
assuming that 1)~the (repulsive) interactions between atoms consist
exclusively of two-body low-energy collisions that can be described
by a delta-function pseudopotential~\cite{HuangYang} of strength~$g$
(related to its dimensionless counterpart~$\gammat$ through
$g\equiv\hwt\xO^{2}\gammat$), that 2)~the many-body field
operator~$\Psi$ can be decomposed via
\begin{equation}
	\Psi = \langle\Psi\rangle + \psit \equiv \Phit + \psit,
\label{eq:decomp}
\end{equation}
where the ensemble average $\Phit$ is a real macroscopic wavefunction
that describes the condensate (reflecting the imposition of
macroscopic long-range order on the condensed system), and that
3)~products of noncondensate operators can be simplified using a
finite-temperature version of Wick's theorem~\cite{FetWal}.  If we
insert~(\ref{eq:decomp}) into the many-body grand-canonical
Hamiltonian
\begin{equation}
	\tilde H = \int\dV{2}{x}\Psid(\Lambdat
	+ \frac{\gammat}{2}
	\Psid\Psid\Psi)\Psi,
\label{eq:HamN}
\end{equation}
where $\Lambdat = -\lapt+x^2-\mut$, and neglect anomalous averages via
the Popov approximation~\cite{GPS}, we obtain an expression that can
be diagonalized and yields an infinite set of coupled differential
equations.

On one hand, the macroscopic wavefunction mentioned above is the
square root of the dimensionless condensate density~$n_0$ and obeys
the generalized Gross-Pitaevski\u\i{} equation,
\begin{equation}
	\Lambdat\Phit+\gammat(\ennt_{0}+{2\ennt'})\Phit = 0,
\label{eq:GP}
\end{equation}
where $\enet'\equiv\enet-\enet_0$~is the noncondensate density.  The
factor of~$2$ in~(\ref{eq:GP}) and hereafter is a consequence of the
direct (Hartree) and exchange (Fock) terms being identical, which
follows from the fact that the delta-function interaction that we are
considering has zero range~\cite{Griffin96}.  (The term involving the
condensate does not include that factor; this is a consequence of the
restricted grand-canonical ensemble that we are using.  See the end
of the next section.)

The noncondensate, in turn, is described by an infinite number of
pairs of functions that obey~\cite{Fetter72}
\begin{equation}
	\left(\begin{array}{cc}
		\eLe & -\gammat\enet_0 \\ -\gammat\enet_0 & \eLe
	\end{array}\right)
	\left(\begin{array}{cc}
		u_j \\ v_j
	\end{array}\right) = \epsit_j\,
	\left(\begin{array}{cc}
		u_j \\ -v_j
	\end{array}\right)
\label{eq:Exact}
\end{equation}
and which generate the noncondensate density via
\begin{equation}
\enet'(\xx) = \sum_j\,\big(\big(
	\big|u_j\big|^2+\big|v_j\big|^2\big)
		f_j + \big|v_j\big|^2\big),
\label{eq:ExDens}
\end{equation}
where we have introduced the Bose-Einstein distribution factor
$f_j\equiv(e^{\epsit_j/t}-1)^{-1}$, which appears when the free energy
of the system is minimized, and the dimensionless
temperature~$t\equiv\kB T/\hwt$.  The last term of~(\ref{eq:ExDens})
describes the zero-temperature depletion of the condensate, which
accounts for less than 1\%{} of the particles and is therefore
negligible~\cite{GPS}.  This self-consistent set of equations is
closed, and the chemical potential found, by imposing that the total
number of particles remain fixed:
\begin{equation}
	N = N_0 + N' = \int\dV{2}{x} \enet(\xx) = 
		\int\dV{2}{x} 
		\left(\enet_0+\enet'\right).
\label{eq:Conserv}
\end{equation}

At temperatures such that $\kB T\gg\hw$, one can use the semiclassical
(WKB) approximation~\cite{VdB} that results from taking $u_j\approx
u(\xx)e^{i\phi}$ and $v_j\approx v(\xx)e^{i\phi}$.  The phase common
to both defines a quasiparticle momentum through $\kk = \gradt\phi$;
$u$,~$v$, and~$\kk$ vary sufficiently slowly with~$\xx$ that their
spatial derivatives can be neglected~\cite{GPS}.  The distribution
factor is now~$f_j\approx f(\xx,\kk)$, infinite sums are transformed
into momentum integrals---which in two dimensions can be solved in
closed form~\cite{mulShort}---and the equations become algebraic,
yielding the Bogoliubov energy spectrum~\cite{GPS}
\begin{eqnarray}
	\eHFB(\kk,\xx) =
		\sqrt{(\kappa^2 + x^2 + 2\gammat\enet - \mut)^2
		-\gammat^2\enet_0^2} 
\label{eq:BogSpect}
\end{eqnarray}
and the following integral expression for the noncondensate density:
\begin{eqnarray}
	\enet'(\xx) 
	\EQQ \hphantom{-}\frac{t}{4\pi}\,
	\int_{
		(\Zeff^2-\gammat^2\enet_0^2)^{1/2}/t
	}^\infty
		\frac{d\xi}{e^\xi - 1}
	\nonumber \\
	\EQQ -\frac{t}{4\pi}\,
	\log\big(1-e^{-(\sqrt{\Zeff^2-\gammat^2\enet_0^2})/t}
	\big).
\label{eq:HFB}
\end{eqnarray}
At high enough temperatures there is no condensate, and the density on
the left-hand side of~(\ref{eq:HFB}) is just the total density.  We
thus have to solve a single self-consistent equation,
\begin{equation}
	\enet'(\xx)\to\enet(\xx) = -\frac{t}{4\pi}\,
	\log\big(1-e^{-\Zeff/t}
	\big);
\label{eq:Brandon}
\end{equation}
the chemical potential is once again calculated by requiring $N$~to be
fixed.  Equation~(\ref{eq:Brandon}) has been found by the authors of
\Ref{bhaduri} to be soluble at \emph{all temperatures}, a result that
we confirm in the present work.  Reference~\citen{mulShort} had
mistakenly concluded that it was impossible to
solve~(\ref{eq:Brandon}) below a certain temperature.

The Hartree-Fock approximation~\cite{GSL,GPS,HuseSiggia}
amounts to neglecting the~$v_j$ in~(\ref{eq:Exact})
and, when combined with the semiclassical treatment, results in the
disappearance of the last term within the square root of
Eq.~(\ref{eq:BogSpect}).  The energy spectrum now becomes
\begin{equation}
	\eHF\equiv\epsit = \kappa^2+x^2+2\gammat\enet-\mut
\end{equation}
and the noncondensate density turns into
\begin{equation}
	\enet'(\xx) = -\frac{t}{4\pi}\,
	\log\big(1-e^{-\Zeff/t}\big),
\label{eq:HF}
\end{equation}
an expression that differs from~(\ref{eq:Brandon}) in that the
left-hand side corresponds only to the noncondensate density.  The
three-dimensional version of this equation, coupled with~(\ref{eq:GP})
and~(\ref{eq:Conserv}), has been frequently used in the literature to
study the three-dimensional gas.  Reference~\citen{HKN}, for example,
exhibits a detailed comparison of its predictions to those of Monte
Carlo simulations and finds excellent agreement between the two.
\begin{figure}[t]
	\begin{center}
		\resizebox{3.3in}{!}
	{\includegraphics{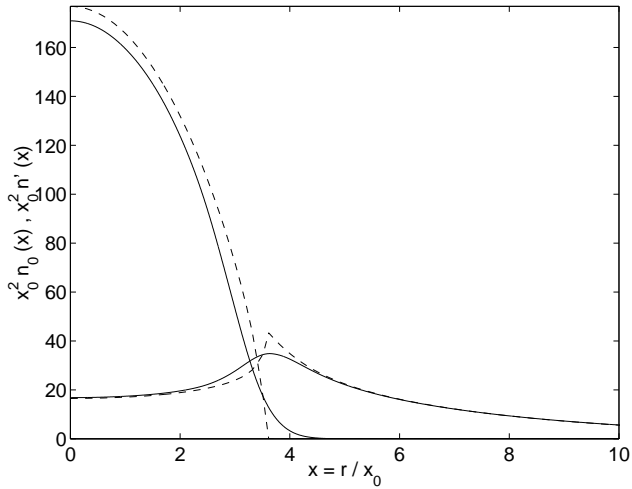}} 
	\end{center}
	\caption{Condensate and noncondensate density profiles of a
		two-dimensional gas with $N = 10^4$ and $\gammat =
		0.1$ at $T = 0.7T_c$, where $T_c$ is the ideal-gas
		transition temperature.  The Goldman-Silvera-Leggett
		model (dashed lines) treats the condensate in the
		Thomas-Fermi limit~(\ref{eq:TF}), while the 
		Hartree-Fock model (full lines) uses the full 
		Gross-Pitaevski\u\i{} equation~(\ref{eq:GP}).
		Both models use \Eq{HF} to describe the
		noncondensate.}
	\label{fig:Solutions}
\end{figure}

In two dimensions, the Hartree-Fock equations have been solved without
resorting to the WKB approximation~\cite{kim2}.  The authors of this
reference succeeded in finding self-consistent density profiles and
used them to study the temperature dependence of the condensate
fraction.  Our results~\cite{Ibiza}, which do take advantage of the
semiclassical approximation, agree quite well with theirs (see
\Fig{Solutions}).

\subsection{The Thermodynamic Limit} \label{sec:TDL}

When $N$~is large we can neglect the kinetic energy of the system,
which in~2D can be shown to be of order $1/N$~\cite{mulShort}, and
obtain the Thomas-Fermi approximation~\cite{BaymPeth}
\begin{equation}
	\gammat\ennt_0 = (\mut - x^2 - 2\gammat\ennt')
		\Theta(\mut - x^2 - 2\gammat\ennt'),
\label{eq:TF}
\end{equation}
where $\Theta(x)$~is the Heaviside step function, introduced to ensure
that the density profile is everywhere real and positive.  Also, since
the thermodynamic limit requires that~$\omega\to0$, the
WKB~approximation becomes rigorous and can be used with confidence.
Thus we can insert expression~(\ref{eq:TF}) in the Bogoliubov energy
spectrum~(\ref{eq:BogSpect}) and show that the latter reduces to
\begin{equation}
	\epsit_{\mbox{\tiny HFB}}(\kk,\xx) \approx
		\kappa\sqrt{\kappa^2 + 2\gammat\enet_0}
		\approx\kappa\sqrt{2\gammat\enet_0}
\end{equation}
for small quasimomenta.  The fact that it is linear clearly shows us
that, in this approximation, the low end of the energy spectrum
corresponds to phononlike quasiparticles.  Now, if we
introduce~(\ref{eq:TF}) into the noncondensate density~(\ref{eq:HFB}),
we can see that the argument of the logarithm vanishes at all
temperatures, making the density diverge at every point in space.  The
first line of~(\ref{eq:HFB}) shows that the divergence in the integral
is caused at its lower limit; this restates the conclusion arrived at
in \Ref{mulShort}: low-energy phonons destabilize the condensate in
the two-dimensional thermodynamic limit when the noncondensate
quasiparticles obey the Bogoliubov spectrum.

In the Hartree-Fock approximation, on the other hand, the energy
spectrum tends in this limit to $\epsit\approx\kappa^2+\gammat\enet_0$
and predicts single-particle excitations whose minimum energy
is~$\gammat\enet_0$; this can be interpreted equivalently by assigning
a minimum value~$\kappa_c^2=\gammat\enet_0$ for the excitation
quasimomentum.  This cutoff is consistent with those proposed
in the past~\cite{fishhoh,kagan87,kagan00,prokofio,popovbook} and removes
the infrared singularity in the HFB equations.

This momentum cutoff is robust enough that it enables one to carry out
Hartree-Fock calculations even in the thermodynamic limit: in fact, as
was first found in \Ref{GSL}, the introduction of this limit actually
simplifies the calculations, and it is possible to find
self-consistent solutions by simultaneously treating the noncondensate
in the Hartree-Fock approximation and the condensate in the
Thomas-Fermi limit~\cite{kim2,Ibiza}.  This model cannot provide
realistic density profiles at every point in space, since the
condensate density~(\ref{eq:TF}) has a discontinuous derivative at its
edge, but predicts quite reasonable results outside of this region, as
can be seen in \Fig{Solutions}.

\subsection{The Free Energy} \label{sec:Energy}

We have seen that in 2D the mean-field-theory BEC equations admit
solutions both with and without a condensate.  The unphysical solution
should be that with the highest free energy, since equilibrium at
finite temperatures occurs when the grand potential attains a minimum;
in fact, the Bose-Einstein distribution factor in \Eq{ExDens} comes
from minimizing this quantity~\cite{HuseSiggia,deGennes}, which in our
dimensionless units adopts the form
\begin{equation}
	\Omega = (U - \mu N - TS) / \hwt
	\equiv
	\int\dV{2}{x}(\Upsilon - \mut\enet - t\Sigma).
\label{eq:FREEdim}
\end{equation}
At this point we have to keep in mind that the grand-canonical free
energy is a function of~$\mut$, not of~$N$; in order to make a
meaningful comparison of these energies at the same~$N$, then, we have
to compare the Helmholtz free energies, given
by~$\tilde{A}=\Omega+\mut N$.  The expressions given below have all
been derived in the grand-canonical ensemble and thus contain the
chemical potential; we will eliminate this dependence on $\mut$ by
adding $\mut N$ to the expressions we obtain.

In the Hartree-Fock approximation, the grand-canonical energy density
of the system is given by~\cite{GPS}
\begin{equation}
	\Upsilon-\mut\enet = \Phit(\Lambdat + 
		{\gammat\over2}\enet_0)\Phit
		+ \frac{1}{(2\pi)^2} 
		\int\frac{d^2\!\kappa\,\epsit}{e^{\epsit/t}-1}
		- \gammat \enet^{\prime 2};
\label{eq:EDens1}
\end{equation}
the first term corresponds to the condensate energy, while the second
one is the sum, weighted by the Bose-Einstein distribution, of the
energies of the excited states; this last expression includes an extra
term~$\gammat\enet^{\prime 2}$ that has to be subtracted explicitly,
as is usually the case in Hartree-Fock calculations~\cite{FetWal}.  We
can simplify~(\ref{eq:EDens1}) further by invoking \Eq{GP}:
\begin{equation}
	\Upsilon-\mut\enet = -\gammat\enet^2 +
		{\gammat\over2}\enet_0^2
		+ \frac{1}{(2\pi)^2} 
		\int\frac{d^2\!\kappa\,\epsit}{e^{\epsit/t}-1}.
\label{eq:EDens}
\end{equation}

The entropy of the system can be found from the combinatorial
expression~\cite{LanLif,FetWal}
\begin{equation}
	S = -\kB\sum_i (f_i\log f_i - (f_i + 1)\log(f_i + 1)),
\label{eq:Entropy}
\end{equation}
which in the WKB and Hartree-Fock approximations yields the
entropy density~\cite{GPS}
\begin{equation}
	t\Sigma = \frac{1}{(2\pi)^2}\int\frac{d^2\!\kappa\,\epsit}
		{e^{\epsit/t}-1}
		-\frac{t}{(2\pi)^2}\int\dV{2}{\kappa}
		\log(1-e^{\epsit/t}).
\label{eq:SDens}
\end{equation}
The first term in~(\ref{eq:SDens}) cancels with the last one
in~(\ref{eq:EDens}), and the other term can be integrated in closed
form, yielding
%
%
%
\begin{eqnarray}
	\tilde{A}_{\mathrm{c}} \EQQ \mut N -
		\int\dV{2}{x} \big( \gammat
		(\enet^2-{1\over2}\enet_0^2) 
			\qquad\qquad\qquad \nonumber \\
		\EQQh
			\qquad\qquad\qquad
		 +\, \frac{t^2}{4\pi}\, \gee{2}{-\Zeff/t}\big)
\label{eq:FreeC}
\end{eqnarray}
for the free energy of the condensed solution.  (Here we have
introduced the Bose-Einstein integral~$g_\sigma(x)\equiv
\sum_{k=1}^\infty x^k/k^\sigma$.  Note that $g_1(x)=-\log(1-x)$, a
function that repeatedly appears above.)  As~$T\to0$, this expression
reduces to the correct value in the homogeneous case~\cite{BagKlep}.
In the thermodynamic limit, we can find an expression for the free
energy per particle at~$T=0$ using the Thomas-Fermi density profile:
\begin{equation}
	\frac{\tilde{A}_{\mbox{\tiny TF}}}{N} = \frac{2}{3}\,
		\sqrt{\frac{2}{\pi}}\,\sqrt{N\gammat}.
\label{eq:FreeTF}
\end{equation}
The free energy of the uncondensed solution is found by simply
crossing out the condensate density in~(\ref{eq:FreeC}):
\begin{equation}
	\tilde{A}_{\mathrm{u}} = \mut N-\int\dV{2}{x} 
	\left( \gammat\enet^2
	+ \frac{t^2}{4\pi}\, \gee{2}{-\Zeff/t}\right).
\label{eq:FreeNoC}
\end{equation}
Equations (\ref{eq:FreeC})~and~(\ref{eq:FreeNoC}), of which
more-general versions are derived in \Ref{HoubStoof} by a different
method~\cite{HuseSiggia}, can be easily seen to become identical at
temperatures high enough that the condensate density can be neglected.
On the other hand, their low-temperature limits differ, since
\Eq{FreeNoC} tends to 
%
%
a higher value than that attained by~(\ref{eq:FreeC}).  This can be
traced back to the fact that the grand-canonical ensemble has to be
changed in order for it to correctly describe the particle-number
fluctuations at low temperatures~\cite{HoubStoof,BergemLi}.  Now, a 2D
Bose system has a condensate at least at~$T=0$, so the two free
energies should coincide there; however, we must keep in mind that the
semiclassical approximation used to derive \Eq{FreeNoC} requires
that~$\kB T\gg\hw$; thus the method we are using \emph{cannot}
describe the appearance of the zero-temperature condensate in the
uncondensed solution.  At the end of the next section we will study
further consequences of this distinction.

\section{Numerical Methods and Results} \label{sec:Res}

There is a time-honored prescription~\cite{Griffin96,GPS} for finding
the self-consistent solution of the Hartree-Fock equations in the
presence of a condensate: Initially, we assume that only the
condensate is present and solve the Gross-Pitaevski{\u\i}
equation~(\ref{eq:GP}) for~$\enet'=0$.  The wavefunction and
eigenvalue that result are fed into the Hartree-Fock expression for
the density~(\ref{eq:HF}), which, when integrated over all space,
yields also a value for the noncondensate fraction~$N'$; one can then
readjust the condensate fraction and solve the Gross-Pitaevski{\u\i}
equation that results.  The process is then iterated until the
chemical potential and the particle fractions stop changing.

By far the most difficult part of this process is the solution of the
nonlinear eigenvalue problem~(\ref{eq:GP}).  Different methods exist
in the literature; we have obtained identical results by solving it as
an initial value problem~\cite{edw} and, much more efficiently, by
employing the method of spline minimization~\cite{Krauth,HKN}.  This
method uses the fact that \Eq{GP} is the Euler-Lagrange equation that
minimizes the functional
\begin{equation}
	J[\Phit] = \int\dV{2}{x}
		\bigl[ 
		(\gradt\Phit)^2
		+ \Phit(x^2+2\gammat\enet(\xx))\Phit
		+ \frac{\gammat}{2}\,\Phit^4
		\bigr].
\label{eq:Functional}
\end{equation}
After setting a small, nonuniform grid of fixed abscissas that
represent the c\oo/rdinate~$x$, we take the corresponding ordinates,
which represent~$\Phit$, as the parameters to be varied until
(\ref{eq:Functional})~attains its smallest possible value.  Since we
only have information about the value of the function at a discrete
set of points, in order to calculate the necessary derivatives and to
integrate we perform a cubic-spline interpolation; the integral is
found using ten-point Gauss-Legendre quadrature.  The minimization is
carried out using the Nelder-Mead method.

We checked our code by comparing its predictions in three dimensions
to previously published results.  For the case studied in \Ref{HKN},
the condensate fractions we found differed from those in the paper by
less than one part in $10^4$.  The code also reproduced previously
known results for the ideal gas, including finite-size
effects~\cite{mulLong}.
\begin{figure}[t]
	\begin{center} 
		\resizebox{3.3in}{!}
	{\includegraphics{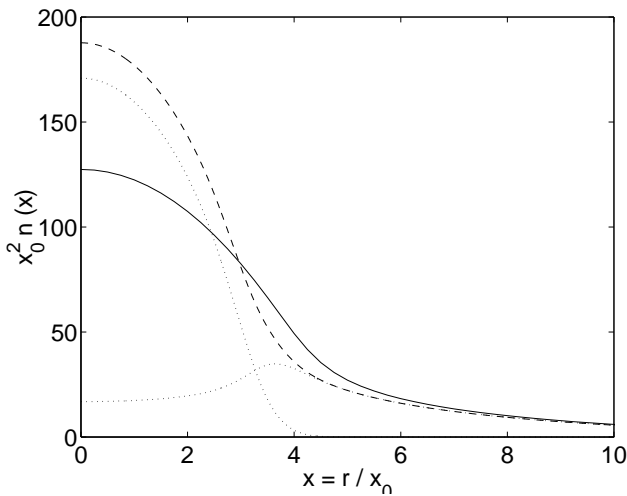}}
	\end{center} 
	\caption{Total density profiles of a two-dimensional gas with 
		the same parameters as in Fig.~\ref{fig:Solutions}.  In 
		this case we exhibit both the uncondensed (full 
		line) and the condensed (dashed line) solutions; 
		the latter we have broken once again into its
		condensate and noncondensate parts (dotted lines).
		}
		\label{fig:NoCond}
\end{figure}

Figure~\ref{fig:Solutions}, already discussed above, shows one of the
solutions that we have found using the Hartree-Fock approximation.
The gas has~$N=10^4$ atoms; the coupling constant~$\gammat$ has been
chosen so that the system has approximately the same radius as the
three-dimensional gas studied in \Ref{HKN}, where parameters
resembling those of the original JILA trap~\cite{Wieman} are used.
The system is shown at $T=0.7T_c$, where $T_c$ is the condensation
temperature for the ideal gas.  We have also found solutions for the
Goldman-Silvera-Leggett model, which corresponds to the large-$N$
limit of the GP equation.  This was done by treating the problem as a
simultaneous system of nonlinear equations on a uniform grid and
solving it with a least-squares method.

When~$N=10^4$, as in \Fig{Solutions}, it is not possible to find a
self-consistent solution for the condensed equations
beyond~$T\approx0.8T_c$, since $\eHF$~becomes negative; this had
already been noted in \Ref{HKN} for three dimensions, where it was
interpreted as a finite-size effect, and occurs at even lower
temperatures for~2D.  The limitation becomes more severe as
$N$~increases: for~$N=10^6$ we cannot find solutions
beyond~$T\approx0.5T_c$.  The authors of \Ref{kim2} report predictions
at temperatures very close to the transition by using a finite-size
correction to the chemical potential~\cite{nearTc}.  Our inability to
work above certain temperatures might be a consequence of using the
semiclassical approximation, though we point out that the Popov
approximation is expected to break down close to the transition
temperature~\cite{GPS,RMPBEC}.

We also found self-consistent three-dimensional solutions using the
more general Bogoliubov spectrum by applying the same method and using
the Hartree-Fock solutions as a starting point for the iteration.  We
find that these solutions exhibit an enhanced depletion of the
condensate, in agreement with those found by other
authors~\cite{nearTc}.  It was impossible, however, to find this kind
of solution in two dimensions, even for systems with $N$~as low
as~$100$: after a few iterations, the chemical potential became too
large, the Bogoliubov energies became imaginary, and the noncondensate
density diverged.  When the possibility of phononlike excitations is
allowed, then, the condensate is destabilized, just as had been found
in the Thomas-Fermi limit~\cite{mulShort}.

The uncondensed case, on the other hand, gives us solutions in these
conditions, and to it we now turn.  Equation~(\ref{eq:Brandon}) along
with the particle-conservation condition~(\ref{eq:Conserv}) for all
temperatures is most easily solved by rewriting~(\ref{eq:Brandon}) as
\begin{equation}
	Ze^{-x^2/t} = 2\,e^{-(\pi-\gammat)\nu(\xx)}\sinh\pi\nu(\xx)
\label{eq:Brand1}
\end{equation}
where have introduced the fugacity~$Z=e^{\mut/t}$
and~$\nu(\xx)\equiv2\enet(\xx)/t$.  Given $Z$, $t$,~and~$\gammat$ it
is possible to find $\nu$~at every point using a standard root-finding
algorithm.  One then wants to find the value of~$Z$ such that the
total number of particles is~$N$.  It is better, however, to write
\Eq{Brand1} at the origin,
\begin{equation}
	Z = 2\,e^{-(\pi-\gammat)\nu_0}\sinh\pi\nu_0,
\label{eq:Brand2}
\end{equation}
eliminate $Z$ between (\ref{eq:Brand1}) and (\ref{eq:Brand2}), and
solve for~$\nu_0=\nu(0)$, the density at the center of the trap,
using the same root finder.

In \Fig{NoCond} we show the (Hartree-Fock) condensed and uncondensed
solutions that we have obtained.  They are similar in shape and
exhibit identical behavior for large~$x$.  The uncondensed solution
has a lower value at the origin and predicts a wider radial density
profile.

\begin{figure}[t]
	\begin{center} 
		\resizebox{3.3in}{!}
	{\rotatebox{90}
	{\includegraphics{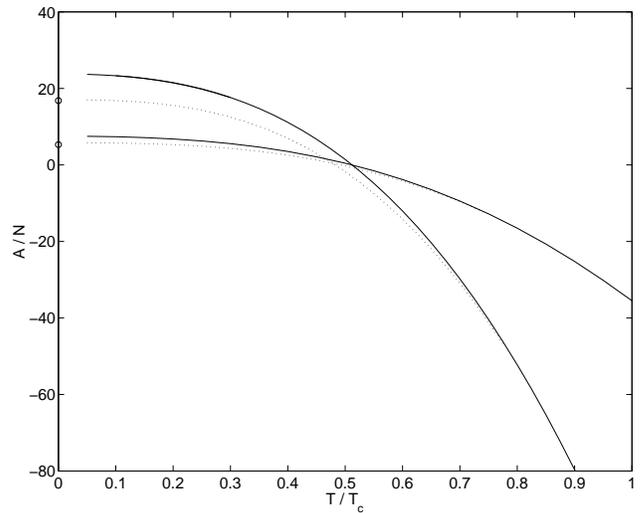}}}
	\end{center} 
	\caption{Free energies per particle for the condensed (dashed
		line) and uncondensed (full line) solutions.  The curves 
		with higher value at~$T=0$ correspond to $N=10^4$~atoms, 
		while those with the lower value at~$T=0$
		correspond to~$N=10^4$.  The coupling constant
		is~$\gammat=0.1$ in both cases.  The open circles on
		the vertical axis are the Thomas-Fermi predictions
		given by \Eq{FreeTF} for~$T=0$.  The free energies seem 
		to be converging in the high-temperature limit; 
		the low-temperature limit, on the other hand, shows
		the discrepancy that results from restricting
		condensate fluctuations (see discussion after \Eq{FreeNoC}).
		Note that~$T_c$, the transition temperature for the
		ideal 2D~trapped gas, depends on~$N$; thus the $x$~axis
		for each~$N$ represents a different actual temperature.
		}
		\label{fig:Free}
\end{figure}

We also calculated the free energy corresponding to each solution; the
results are shown in \Fig{Free}, which shows $\tilde{A}/N$ as a function of
temperature for both cases when $N=10^3$~and~$N=10^4$.  The free
energies appear to coincide at high temperatures; this was to be
expected, since Eqs.~(\ref{eq:FreeC})~and~(\ref{eq:FreeNoC}) become
identical at temperatures high enough for the condensate
density~$\enet_0$ to be neglected.  As for the low-temperature limit,
we have already noted that the free energy of the condensed solution
tends to the value predicted by the zero-temperature Thomas-Fermi
limit, while that of the uncondensed solution tends to a higher value.
This value can actually be calculated: it is easy to
show~\cite{bhaduri} that the low-$T$ limit of \Eq{Brand1} is
\begin{equation}
	2\gammat n(\xx) = \mut - x^2,
\label{eq:BhadTF}
\end{equation}
the Thomas-Fermi limit but with~$\gammat$ replaced by~$2\gammat$; a
larger interaction strength, as we can see from~(\ref{eq:FreeTF}),
implies a higher free energy.  Interestingly, when we compare a given
Hartree-Fock solution to an uncondensed solution with half the
interaction strength, we find that the density profiles become very
similar (and are indistinguishable at $T=0$), while the free energies
coincide almost exactly for a wide range of temperatures; attractive
as this possibility might be, however, it has a serious flaw: the free
energies start to differ as the temperature increases, where they
should coincide by definition.  This tells us that the factor of~$2$,
which turns out to be the same one discussed after
Eqs.~(\ref{eq:GP})~and~(\ref{eq:FreeNoC}), has to be retained;
the WKB approximate expression for the uncondensed state is valid
\emph{only} for~$\kB T\gg\hw$, so dropping the factor of~$2$ in
order to match the $T=0$~condensate is invalid.
(The authors of~\Ref{bhaduri}, in fact, omit this factor from their
paper, although they address this question in a subsequent
publication~\cite{brandon}).  It is in fact this factor that
guarantees that the uncondensed solution has a higher free energy than
the condensed one at all temperatures; this, despite our
inability to find condensed solutions using the Bogoliubov energy
spectrum, leads us to conclude that, at least at this level of
approximation, the uncondensed solution is unphysical and the
two-dimensional finite trapped system will exhibit some sort of 
condensation at finite temperature.

\section{Conclusion} \label{sec:Conc}

We have found solutions to the two-dimensional HFB equations in the
Hartree-Fock approximation, both for finite numbers of atoms and in
the thermodynamic limit; still, when we try to go beyond this scheme
and consider the whole Bogoliubov excitation spectrum, the low end of
which is described by phonons, we are unable to find self-consistent
solutions for finite---even low---values of~$N$.  We have seen that it
is possible to describe the system as having no condensate at all, but
these solutions, at least for the parameter combinations that we have
studied, have a higher free energy than their partly-condensed
counterparts; this leads us to conclude that the 2D system will have
some kind of condensate at low enough temperatures, insofar as the
Hartree-Fock approach can successfully describe the quasicondensate
that we expect to find in a finite system.
%
%
%

One could argue that the whole mean-field approach we have adopted is
wrong in two dimensions, since a condensation into a single state is
being assumed from the start.  However, an alternative
treatment~\cite{Shev92} that does not make this assumption ends up
with equations identical to~(\ref{eq:Exact}), so we are left none the
wiser.  We have also bypassed the fact that the interaction
strength~$g$ is not really constant in~2D, but rather depends
logarithmically on the relative momentum~\cite{Shev91,petrov}; this,
however, has been found to be of little consequence~\cite{brandon}.
Another possibility is that condensation occurs into a band of states,
forming a ``smeared'' or generalized condensate~\cite{Gir,Noz}; this
alternative situation is not ruled out by the standard proof of
Hohenberg's theorem~\cite{MulHT}.

%
%

Monte Carlo simulations \emph{do} seem to predict the presence of a
condensate in two dimensions~\cite{Heinrichs,Pearson}.  Furthermore,
two-dimensional condensates appear to have been produced in the
laboratory~\cite{gorlitz,jakkola}.  Presumably the MC~simulations show
both the effects of a quasicondensation of a finite system and the BKT
transition to the superfluid state.  Our attempt here has been to test
the possibility of representing these Bose effects by a relatively
simple set of HFB or Hartree-Fock equations.  While no solutions can
be found for the HFB set, even for a finite system, the Hartree-Fock
equations do provide a description of a condensed state.  We feel
there is reason to believe that this description should be a fair
representation of the actual situation.

\begin{acknowledgments}
Markus Holzmann lent us the program he wrote for \Ref{HKN} and thus
allowed us to compare our results to those in the reference; we are
grateful to him for that, and to P.~G.~Kevrekidis and E.~Navayazdani
for illuminating conversations.  We also thank the
Department of Astronomy at the University of Massachusetts for
granting us access to their computational facilities.  Finally, we
acknowledge the input given to this paper by two anonymous referees.
\end{acknowledgments}

\bibliography{condbib}

\end{document}